\documentclass[preprint,12pt]{elsarticle}

\usepackage{graphicx}
\usepackage{amsmath,amsfonts,amssymb}

\journal{Materials Letters}

\begin{document}

\begin{frontmatter}



\title{A first-principle study on some quanternary Heusler alloys with 4d and 3d transition metal elements
}


\author[mymainaddress]{Qiang Gao}
\author[mymainaddress]{Huan-Huan Xie}
\author[mymainaddress]{Lei Li}
\author[mymainaddress]{Gang Lei}
\author[mymainaddress]{Ke Wang}
\author[mymainaddress]{Jian-Bo Deng}

\author[mymainaddress]{Xian-Ru Hu\corref{mycorrespondingauthor}}
\cortext[mycorrespondingauthor]{Corresponding author}
\ead{huxianru@lzu.edu.cn}

\address[mymainaddress]{School of Physical Science and Technology, Lanzhou University,
 Lanzhou 730000, People's Republic of China}

\begin{abstract}
In this letter, we investigate the magnetic properties, electronic structures, Slater-Pauling behaviours of some quanternary Heusler alloys with 4d and 3d transition metal elements. The energy levels of the minority-spin electronic band structures for all our calculated alloys are discussed. We have summaried 21 quanternary Heusler alloys. Our calculations indicate that the half metals ZrCoCrSi, ZrCoCrIn, ZrCoFeSi, ZrFeVAl, ZrFeVGa and NbFeCrAl have large spin-flip gaps.

\end{abstract}

\end{frontmatter}


\section{Introduction}
Efficient spin injection from a ferromagnet to a semiconductor is very meaningful for the development of the performance of spintronic devices \cite{1}. Besides, since the prediction of half-metallic  ferromagnetism (HMF) of Heusler alloy NiMnSb by first-principle calculations in 1983 \cite{2}, HMF has attracted great interest. And the HMF Heusler alloys are good candidates for the applications of spintronic devices, as these alloys often have high spin polarization, high Curie temperature and compatible lattice structure. There are many investigations on the half metallic Heusler alloys. Lots of interesting properties of them have been found \cite{10}. For example, the so called LiMgPdSn or Y-type structure \cite{3-1,3-2} Heusler compounds have attracted much attention for their potentional value. This structure Heuslers can be half metals (HM) and spin-gapless semiconductors (SGS). The quanternary Heusler ZrFeTiZ (Z=Al, Si and Ge) and ZrNiTiAl have been reported to be HMF recently \cite{4} with large band gaps and spin-flip gaps. 

Inspired by the above, we study a set of 21 quanternary Heusler alloys with 4d and 3d transition metal elements by using density function theory (DFT) calculations. The plan of our letter is as follows: The next Section 2 gives the computational details; Section 3 presents the discussions of the calculated results; Finally the conclusions are given in Section 4.

\section{Computational details}
The lattice optimization, density of states, magnetic properties and band structure of our calculated quanternary Heusler alloys are calculated by DFT calculations. All our DFT calculations are performed by using the full-potential local-orbital minimum-basis band structure scheme (FPLO) \cite{5-1,5-2} with generalized gradient approximation (GGA)\cite{5-3,5-4,5-5}. For the irreducible Brillouin zone, we use the $k$ meshes of 20$\times$20$\times$20 for all the calculations. The convergence criteria of self-consistent iterations is set to $10^{-6}$ to the density and $10^{-8}$ Hartree to the total energy per formula unit. 

\section{Results and discussions}
In our calculations, $X_1$ is a 4d transition metal element Zr or Nb. $X_2$ and Y are the 3d transition metal elements Co, Cr, Fe, V and Ti, which the atomic number of $X_2$ is bigger than that of Y. Z is a main group element. LiMgPdSn or Y-type structure (space group No.216, F$\bar{4}$3m) have four Wyckoff positions 4a(0, 0, 0), 4b($\frac{1}{2}$,$\frac{1}{2}$,$\frac{1}{2}$), 4c($\frac{1}{4}$,$\frac{1}{4}$,$\frac{1}{4}$) and 4d($\frac{3}{4}$,$\frac{3}{4}$,$\frac{3}{4}$). $X_1$, $X_2$, Y and Z atoms can sit at three possible nonequivalent superstructures: Y-type ($\uppercase\expandafter{\romannumeral1}$) (4a, 4b, 4c, 4d), Y-type ($\uppercase\expandafter{\romannumeral2}$) (4a, 4c, 4b, 4d) and Y-type ($\uppercase\expandafter{\romannumeral3}$) (4a, 4d, 4b, 4c). In order to get the equilibrium structures of the alloys, the geometry optimization is performed in their three different configurations by calculating the total energy as a function of lattice constant. According to the calculated results of total energy at equilibrium lattice constant, we can get that Y-type ($\uppercase\expandafter{\romannumeral1}$) in spin-polarization is the most stable structure for all our calculated       alloys. So only the Y-type ($\uppercase\expandafter{\romannumeral1}$) structure in spin-polarization will be discussed. 

The optimizated lattice constants, band gaps, spin-flip gaps and physical nature are shown in \textbf{Table 1}. As can be seen, there are six SGSs. The detail investigation of these SGSs is in our rprevious paper \cite{6}. The discussions of ZrCoFeSi and ZrCoFeGe are in our another paper \cite{7}. As a contrast， we have also studied ZrFeTiAl, and our results fit well with the discussions in Ref. \cite{4}. In our calculations, there are 14 HMs in total. The 14 HMs have large band gaps.

In \textbf{Figure 1}, the total spin magnetic moment ($M_t$/$\mu_B$) versus the number of total valence electrons in the unit cell is plotted. For the case of positive $M_t$, the spin-down and spin-up are the majority and minority states, while in the the case of negative $M_t$ the opposite occurs. Our results can be classfied along two lines representing two variants of the Slater-Pauling (SP) rules. It can be seen the alloys consisted of Nb only comply with the $M_t=Z_t-24$ SP rule. When  $X_1$ is Zr and Z is Be, Al, Ga or In, the quanternary Heusler alloys comply with the  $M_t=Z_t-18$ SP rule. As for the case that $X_1$ is Zr and Z is Si or Ge, the quanternary Heusler alloys are divided into two parts. One part obeys the $M_t=Z_t-18$ SP rule and the other complies with the $M_t=Z_t-24$ SP rule. Attention should be paid to that Zr and Ti are in the same Group \uppercase\expandafter{\romannumeral4} B, while Nb and V are  in the same  Group \uppercase\expandafter{\romannumeral5} B. So the results of Zr is similar to that of Ti \cite{9}, while the results of Nb is similar to that of V \cite{9}. 

To understand the SP behaviors of these quanternary alloys, we present a schematic representation of the energy levels of the minority-spin band structure for them in \textbf{Figure 2}.  The occupied and unoccupied states locate below and above the Fermi level, respectively. The symbol beside each orbital denotes the corresponding degeneracy while we have introduced the nomenclature of \textbf{Ref. \cite{8}}. The origin of the SP rules for the   Heusler alloys with 4d and 3d transition metal elements is due to the hybridization effects which have been discussed detailedly in our recent paper \cite{6}. The Zr-based alloys which obey the $M_t=Z_t-18$ SP rule are similar to the Sc$_2$- and Ti$_2$- based inverse Heusler alloys \cite{9}. As can be seen in \textbf{Figure 2}, the triple-degeneration $t_{1u}$ states and the double-degeneration $e_u$ states are in such high energy level that both states are empty. And the minority-spin gaps are created between the non-bonding $t_{1u}$ and bonding $t_{2g}$ states. In the schematic, one single state 1$\times$s, two triple-degeneration states 3$\times$p as well as 3$\times$$t_{2g}$ and one double-degeneration state 2$\times$$e_g$ are below the Fermi level. So the number of states below the Fermi level: N$\downarrow$=9. We can directly deduce the number of occupied majority-spin states: $N\uparrow=Z_{tot}-N\downarrow$. Hence the total magnetic moment: $M_{tot}=(N\uparrow-N\downarrow)\mu_B=(Z_{tot}-2\times N\downarrow)\mu_B=(Z_{tot}-18)\mu_B$. Therefore some Zr-based alloys obey the $M_t=Z_t-18$ SP rule.

The case of some Zr- and Nb- based alloys which obey the $M_t=Z_t-24$ SP rule is similar to the discussions of the full-Heusler alloys in Ref \cite{8}. As shown in \textbf{Figure 2}, the $t_{1u}$ states are below the Fermi level, and the $e_u$ states are still above the Fermi level. The $t_{1u}$ and $e_u$ states are occupied and empty respectively. As can be seen, the minority-spin gap is created between the nonbonding $t_{1u}$ and $e_u$ states. Below the Fermi level are the 1$\times$s, 3$\times$p, 2$\times$$e_g$, 3$\times$$t_{2g}$ and 3$\times$$t_{1u}$ states. So N$\downarrow$ is 12. Hence we can directly get that: $M_{tot}=(N\uparrow-N\downarrow)\mu_B=(Z_{tot}-2\times N\downarrow)\mu_B=(Z_{tot}-24)\mu_B$. Therefore some Zr- and Nb- based alloys obey the $M_t=Z_t-24$ SP rule.   

In \textbf{Figure 3}, we show the total density  of states (TDOS) for  ZrCoCrSi, ZrCoCrIn, ZrCoFeSi \cite{7}, ZrFeVAl, ZrFeVGa and NbFeCrAl. For the other HMs, we can get similar conclusions.  As can be seen, the TDOS in the majority-spin channel get through the Fermi level, which exhibits a metallic behaviour. And there is a band gap around the Fermi level in the minority-spin channel, which exhibits a semiconductive behaviour. Around the Fermi level, the energy gap in the minority-spin channel and the crossing over in the majority-spin channel lead to the 100\% spin polarization. So these six alloys are HMs. The spin-flip gap is defined as the minimum gap between the Fermi level and the edge of the band gap in the minority-spin channel. Therefore we can get that the spin-flip gaps are 0.10, 0.22, 0.22 \cite{7}, 0.33, 0.38 and 0.21 eV for ZrCoCrSi, ZrCoCrIn, ZrCoFeSi \cite{}, ZrFeVAl, ZrFeVGa and NbFeCrAl, respectively. Hense they may keep their half-metallicity at room temperature. And they may be good candidates for the future spintronics use.

We also discuss the partial density of states (PDOS) of ZrFeVGa as a representative. \textbf{Figure 4}
 is the PDOS and TDOS. As can be seen, the 3d states of Fe and V make the main contribution to the TDOS in the range of -4 to 2 eV. The 4d states make the most contribution among all the states of Zr in this range. And the 4p states of Ga make the most contribution among all the states of Ga near the Fermi level. It is clear that there is a hybridization between the 3d states of Fe and V around the Fermi level.

\section{Conclusions}
In a word, we study a series of quanternary Heusler alloys with 4d and 3d transition metal elements in this letter. Specificly, we study the electronic structures, magnetic properties and Slater-Pauling behaviours of 21 alloys. The Slater-Pauling behaviours of these alloys have been summaried. The energy levels of them are studied in detail. We focus on six HMs which have large enough spin-flip gaps. The density of states of these HMs are investigated. These HMs may be stable at room temperature because of their large spin-flip gaps. And they may be good candidates for the future  spintronics use.





 \bibliographystyle{elsarticle-num}
 \bibliography{Reference.bib}

\begin{table}[!hbp]
\caption{\label{arttype}}
\begin{tabular*}{\textwidth}{@{}l*{15}{@{\extracolsep{0pt plus12pt}}l}}
\hline
\hline
	X$_1$X$_2$YZ&a$_{opt}$(\AA)&E$_g$(eV)&E$_{sfg}$&Nature\\
	\hline
	ZrCoCrSi&6.033&0.27&0.10&HM\\
ZrCoCrAl&6.251&0.93&0.02&HM\\
ZrCoCrBe\cite{6}&6.013&0.71&0.26&SGS\\
ZrCoCrGa&6.236&0.89&0.06&HM\\
ZrCoCrIn&6.455&0.83&0.22&HM\\
ZrCoFeSi\cite{7}&5.971&0.64&0.22&HM\\
ZrCoFeGe\cite{7}&6.056&--&--&Nearly HM\\
ZrCoFeP\cite{6}&5.944&0.41&0.06&SGS\\
ZrFeCrSi&6.039&0.52&0.26&HM\\
ZrFeCrIn\cite{6}&6.419&0.80&0.06&SGS\\
ZrFeCrGa\cite{6}&6.184&$\downarrow$0.71 $\uparrow$0.02&--&SGS\\
ZrFeVAl&6.248&0.93&0.33&HM\\
ZrFeVGe\cite{6}&6.210&$\downarrow$0.81 $\uparrow$0.41&--&SGS\\
ZrFeVGa&6.237&0.98&0.38&HM\\
ZrCoVIn\cite{6}&6.468&0.98&0.15&SGS \\
ZrFeTiAl\cite{4}&6.31&0.56&--&HM \\
ZrFeTiAl&6.325&0.80&0.22&HM \\                  
NbFeVSi&6.000&--&--&Nearly HM \\
NbFeVGe&6.086&0.18&0.03&HM \\
NbFeCrSi&5.912&0.08&0&HM \\
NbFeCrAl&6.035&0.43&0.21&HM \\
NbCoCrAl&6.026&0.17&0.06&HM \\
\hline
\hline

\end{tabular*}
\end{table}
\textbf{Table Caption}\\
\textbf{ Table 1}: The results of the lattice optimization (a$_{opt}$), band gap (E$_g$), spin-flip gap ($E_{sfg}$) and Physical nature of our calculated quanternary Heusler alloys.\\
\textbf{Figure Captions}\\
\textbf{Figure 1}: Calculated total magnetic moment ($M_t$) as a function of the total number of valence electrons ($Z_t$), in the unit cell for the calculated Heusler alloys.\\
\textbf{Figure 2}: Schematic representation of the energy levels of the minority-spin electronic band structure for all the calculated alloys under study.\\
\textbf{Figure 3}: TDOS for the HMs with enough large spin-flip gaps.\\
\textbf{Figure 4}: PDOS for ZrFeVGa as a representation.
\begin{figure}[htp]
\centering
\includegraphics[scale=0.60]{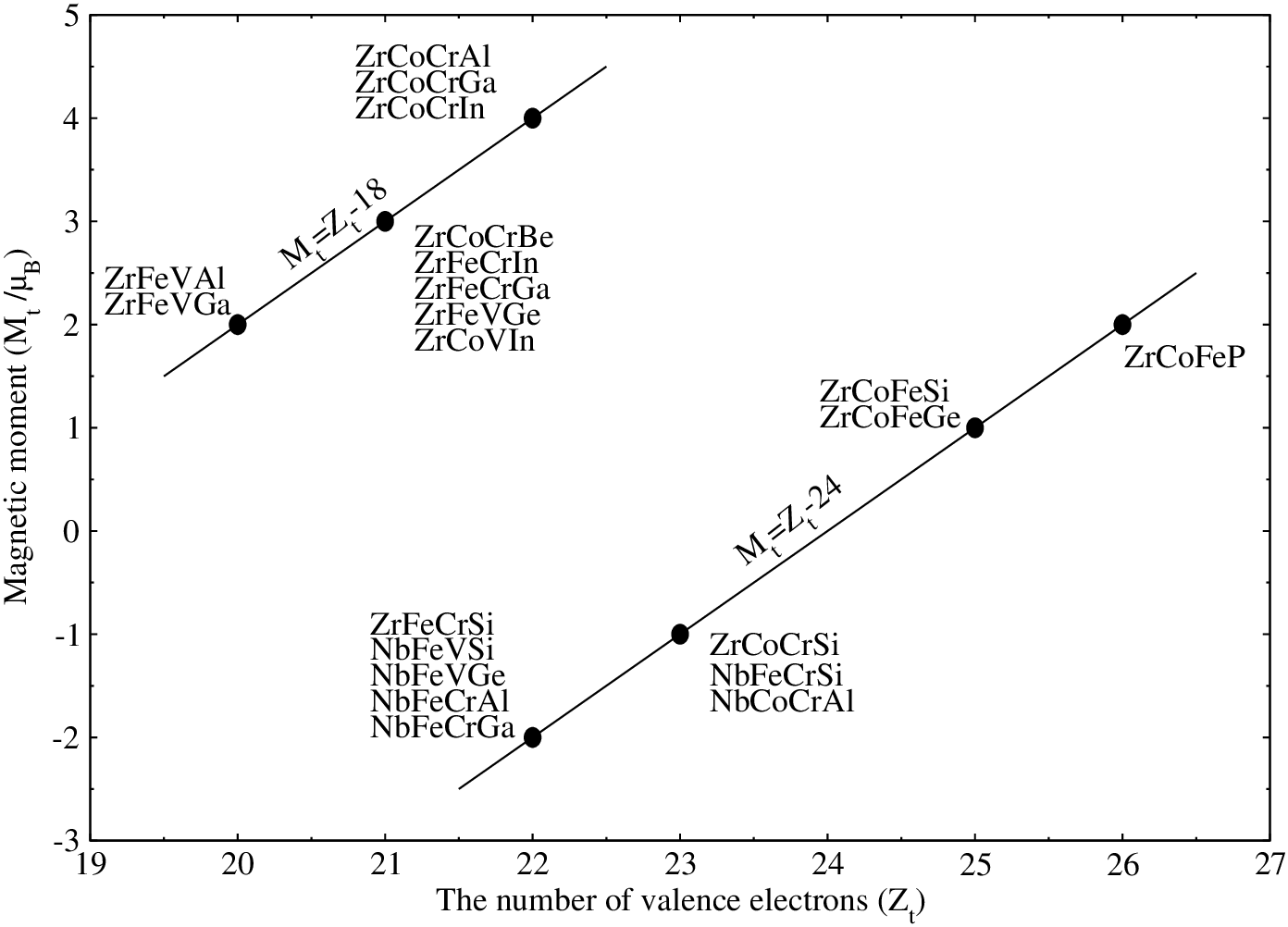}
\caption{}
\label{}
\end{figure}

\begin{figure}[htp]
\centering
\includegraphics[scale=0.60]{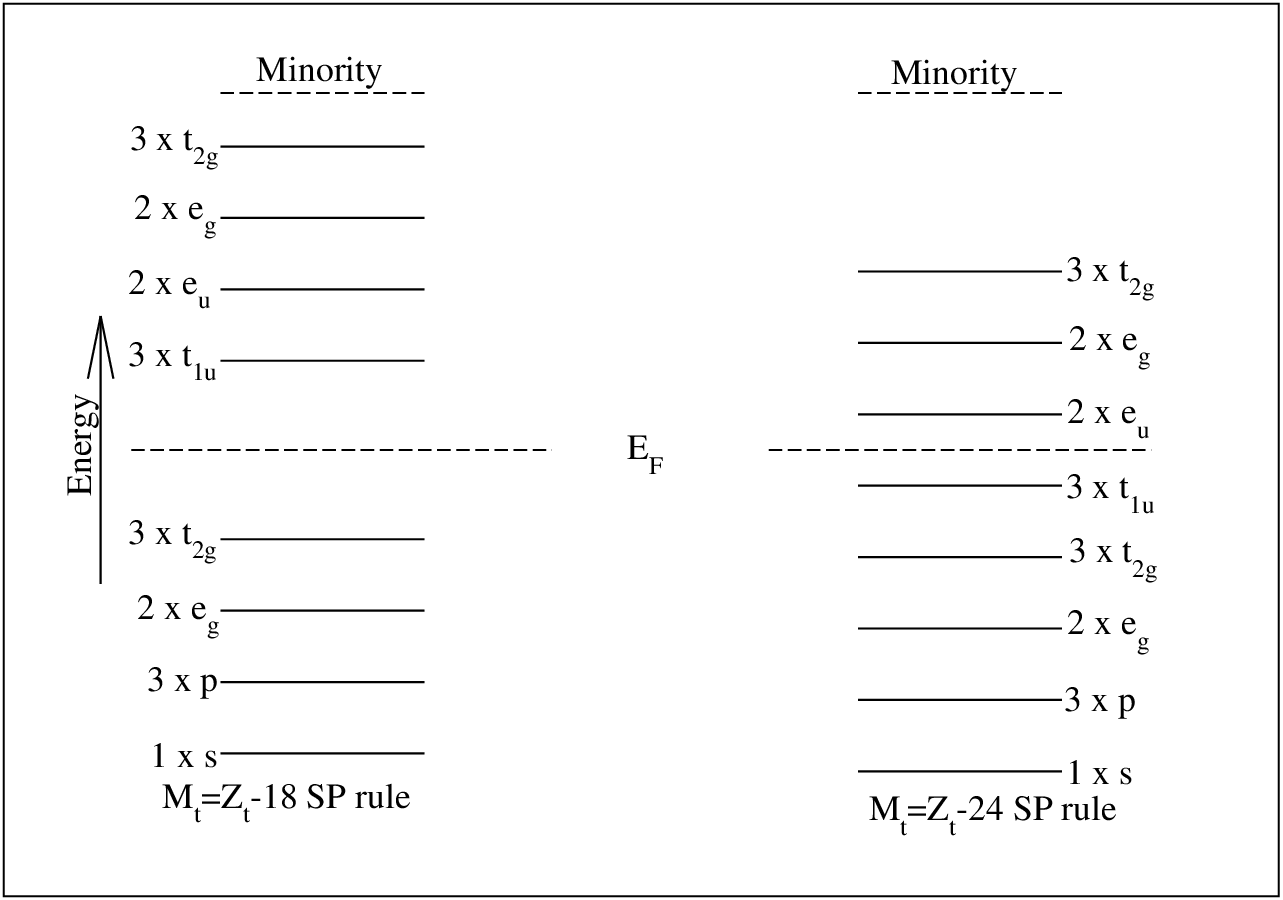}
\caption{}
\label{}
\end{figure}

\begin{figure}[htp]
\centering
\includegraphics[scale=0.60]{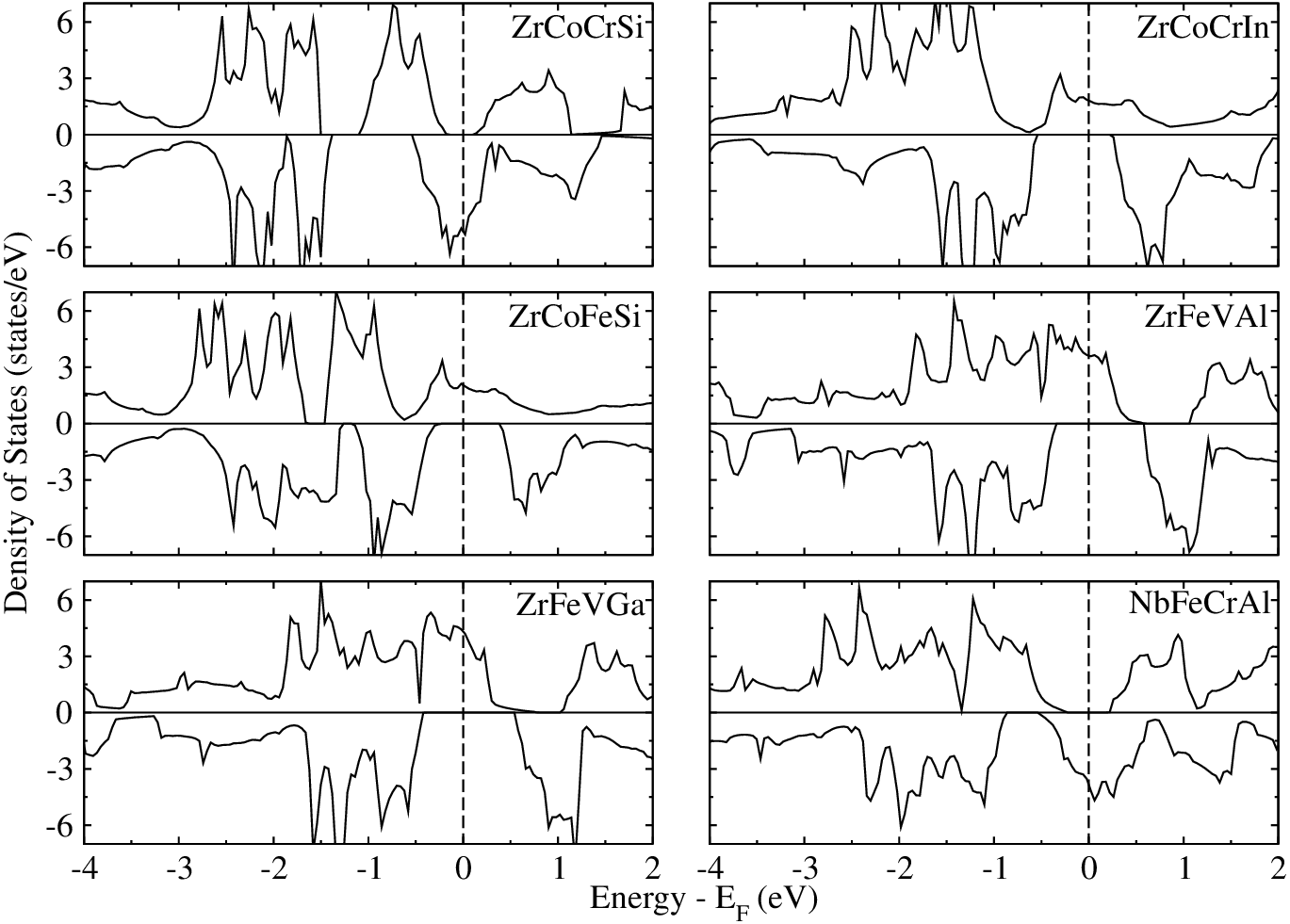}
\caption{}
\label{}
\end{figure}
\begin{figure}[htp]
\centering
\includegraphics[scale=0.60]{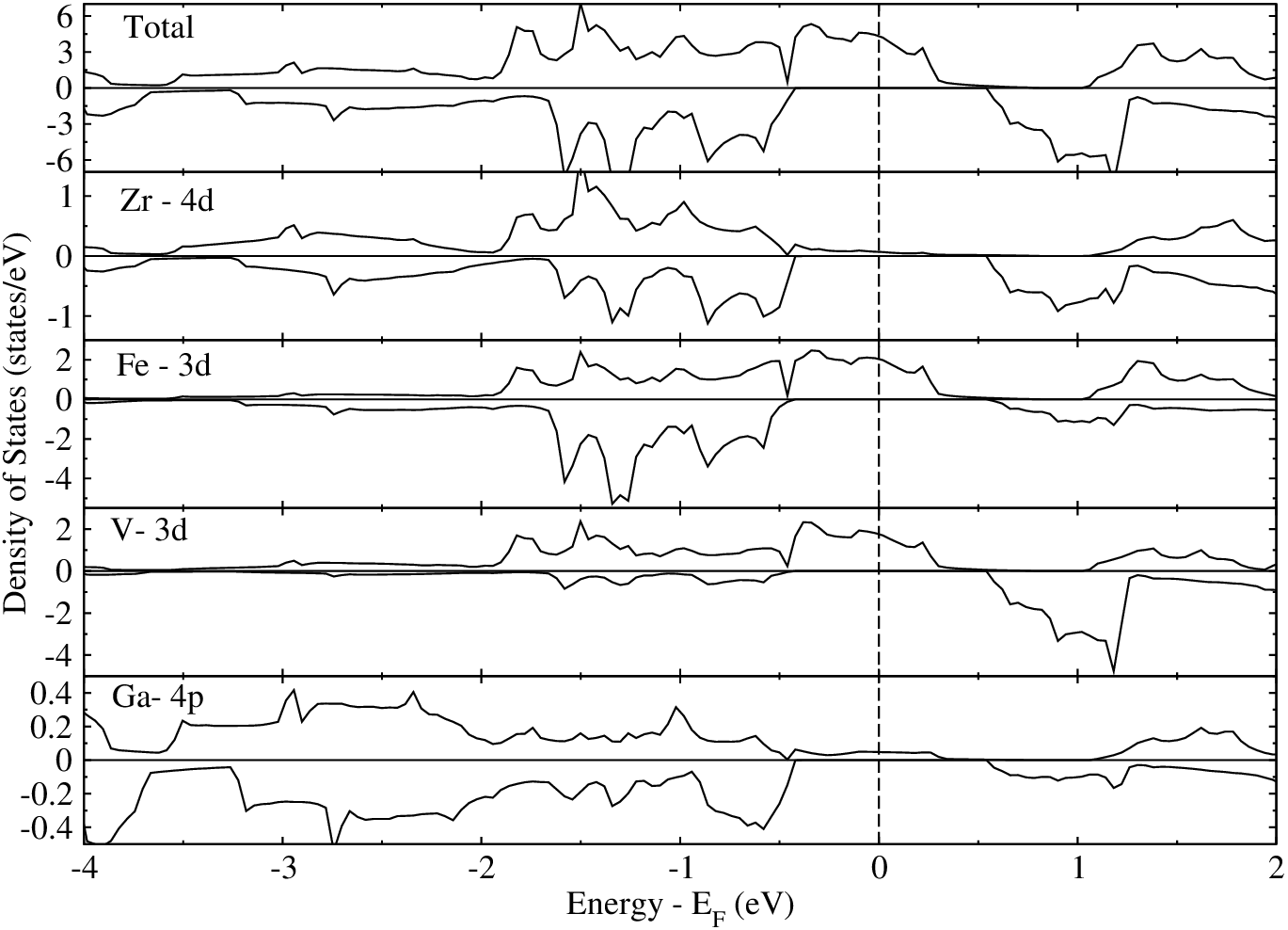}
\caption{}
\label{}
\end{figure}

\end{document}